\documentclass[pdflatex,sn-mathphys-ay]{sn-jnl}

\usepackage{graphicx}%
\usepackage{multirow}%
\usepackage{amsmath,amssymb,amsfonts}%
\usepackage{amsthm}%
\usepackage{mathrsfs}%
\usepackage[title]{appendix}%
\usepackage{xcolor}%
\usepackage{textcomp}%
\usepackage{manyfoot}%
\usepackage{booktabs}%
\usepackage{algorithm}%
\usepackage{algorithmicx}%
\usepackage{algpseudocode}%
\usepackage{listings}%
\usepackage{hyperref}%

\theoremstyle{thmstyleone}%
\theoremstyle{thmstyletwo}%

\theoremstyle{thmstylethree}%

\begin{document}

\title[Article Title]{Young asteroid pair candidates in the Jovian Trojan population}


\author*[1,2]{\fnm{Jind\v{r}ich} \sur{\v{Z}i\v{z}ka}}\email{jinziz@centrum.cz}



\affil*[1]{\orgdiv{Analytical support}, \orgname{Bertin VF Nuclear}, \orgaddress{\street{Svitavsk\'{a} 588}, \city{\v{C}ern\'{a} Hora}, \postcode{CZ 67921}, \state{Czech Republic}, \country{}}}




\abstract{The number of asteroid pairs -- unbound asteroids with a common origin and similar heliocentric orbits -- has increased 
rapidly since their discovery. However, to date, only a small number of asteroid pairs have been identified in the outer main belt and 
especially within the Hilda and Jupiter Trojan populations. Except for a few binaries and collisional families, only one 360~Myr-old 
pair (258656) 2002 ES76 -- (635224) 2013 CC41 is known among the Jupiter Trojans. During our survey, using GPU‑accelerated integrations 
of the Jovian swarms, we identified two new promising pair candidates younger than 2.5 Myr. Pair (264119) Georgeorton--2024~CN most 
likely formed either 427–500~kyr or 527–536~kyr ago, whereas pair (8060) Anius--(542262) 2013 BL has an estimated age of 2367–2421~kyr. 
Our results show that young asteroid pairs can arise even within the sparse population of Jupiter Trojans, opening new questions about 
their formation.}




\maketitle

\section{Introduction}\label{sec1}
An asteroid pair consists of two dynamically unbound asteroids that share nearly identical heliocentric 
orbits, suggesting a common origin. The members of such pairs are thought to be remnants of a binary
component separation \citep{Cuk2007}, asteroid collisions, or YORP--induced rotational fission, which 
is likely the primary mechanism \citep{Vokrouhlicky2008a,Pravec2010}.

Since the first young asteroid pairs were discovered by \cite{Vokrouhlicky2008a}, the number of newly
identified cases has steadily increased, fueling growing interest in the topic. Numerous studies have examined
not only individual cases \citep[e.g.][]{Zizka2016,Vokrouhlicky2017a}, but have also identified additional
young asteroid pairs as part of broader analyses \citep[e.g.][]{Pravec2010}. To date, several exceptionally 
young cases have been identified. A possible comet‑like pair 2019~PR2--2019~QR6 on the order of a few hundred years was 
reported by \cite{Fatka2022}, followed shortly thereafter by another extremely young system (458271) 2010~UM26--2010~RN22, 
only a few decades old \citep{Vokrouhlicky2022a}.

The vast majority of identified asteroid pairs lie interior to the 5:2 mean‑motion resonance with Jupiter
\citep[e.g.][]{Honsova1}, encompassing the Hungaria region and the inner and middle main belt. Their number
decreases with increasing heliocentric distance, consistent with YORP‑induced rotational fission being likely the
dominant formation mechanism. Accordingly, only a few cases are known in the outer main belt and in the Hilda 
and Jupiter Trojan populations, with a single 360 Myr-old Trojan pair identified so far \citep{Holt2020}.

Most recent works by \cite{Kyrylenko2021}, \cite{Kyrylenko2024}, and \cite{Honsova2} have significantly expanded
the set of candidate pairs, whereas \cite{Nesvorny2026} focused on identifying young asteroid families,
within which young asteroid pairs may also be present. Although several exceptionally young asteroid pairs have
now been identified across the main belt, no comparably young systems have been detected among the Jupiter
Trojans. The lack of young pairs among Trojans, still consistent with current expectations, nevertheless motivated 
us to examine the Trojan population in greater detail. In this regard, if a population of Trojan pairs were to 
be discovered and subsequently well understood, together with their formation processes, it could in turn offer 
new insights into the evolution of Trojan binaries.

\section{Candidate search}\label{sec2}
To search for new candidate pairs, we selected all nominal orbits of numbered and multi-opposition objects from the \texttt{AstDyS} database 
(see \url{https://newton.spacedys.com/astdys2/}) at the initial epoch MJD 61200 with $5.0 < a < 5.4$ AU, $e < 0.2$ 
and $i < 40^\circ$, whose osculating mean longitudes lie within $\pm 60^\circ$ of the nominal L4 and L5 centers.
To avoid collisional pairing, we excluded from this set all orbits corresponding to the parent bodies of 
published collisional families listed in \cite{Vokrouhlicky2024} and \cite{Vinogradova2020}. These objects are 
not used in any subsequent dynamical simulations, with 624 Hektor and 617 Patroclus treated only as the 
perturbers in the L4 and L5 swarms, respectively. This left us with 8,668 L4 and 5,449 L5 objects.

\subsection{Candidate search in the 5D space of osculating elements}\label{subsec21}
Using our CUDA‑C code, based on the well‑tested SWIFT‑MVS integrator \citep{Levison1994} and exploiting GPU
parallelism, we propagated all selected orbits backward in time as massless particles under the gravitational 
perturbations of all planets\footnote{When possible, represented by their barycenters.}, Hektor and Patroclus. 
The integrations were performed with a 1‑day time step and carried out separately for the L4 and L5 populations, 
extending to 2.5 Myr.

We used the distance $d$ in the 5D space of osculating orbital elements as a measure of orbital similarity  
to identify unusually close orbits \citep[see][]{Vokrouhlicky2008a}. The metric $d$ is given by 
\begin{equation}
\left(\frac{d}{na}\right)^2 = k_a\left(\frac{\delta{a}}{a}\right)^2 + k_e(\delta{e})^2 + k_i(\delta{\sin{i}})^2 + k_\Omega(\delta{\Omega})^2 + k_\varpi(\delta{\varpi})^2,\label{hcm_5dim_metric}
\end{equation}
where $n$ is the mean motion and $(\delta{a}, \delta{e}, \delta{\sin{i}}, \delta{\varpi}, \delta{\Omega})$ 
is the separation vector between neighboring orbits. Following \cite{Zappala1990a}, we use  
$k_a = 5/4$ and $k_e = k_i = 2$. The weights $k_\Omega$ and $k_\varpi$ are set to $10^{-4}$, keeping the contribution
from differences in $\Omega$ and $\varpi$ small.

In the course of the integration, every 20 days we computed the distance $d$ for all orbital pairs 
and recorded cases with $d < 10$ m s$^{-1}$. By its end, we identified 12,477 L4 and 1,880 L5 candidate 
pairs -- still a substantial number for follow‑up integrations, even when performed on a GPU. To limit the 
sample, we finally selected the 1,600 most frequently recorded cases in each swarm, as these repeated 
detections point to pairs whose components remained close for longer than during a single flyby.

\subsection{Candidate search in heliocentric coordinates}\label{subsec22}
We further examined an alternative strategy for identifying candidate pairs, this time working directly in
heliocentric coordinates. Massless particles placed on the nominal orbits of the pair components, derived from 
best‑fit astrometric observations, follow the most probable orbital solutions. However, gravitational 
and non‑gravitational perturbations may prevent them from meeting within the Hill sphere at a relative velocity below the 
escape velocity of the parent body — conditions we would expect if they represented a real pair. Nevertheless, 
finite orbital accuracy allows the existence of statistically equivalent realizations of the components slightly 
off their nominal orbits (called clones), potentially yielding a convergent solution even if the nominal 
orbits do not.

To lose as few candidate pairs as possible in our preselection, we back‑propagated 5~clones per object, which,
although limited in number, still better reflect the underlying complexity than a simple nominal orbit. Clones 
were randomly selected from the confidence interval of each object's nominal orbit and backtracked under gravitational 
forces and the thermal acceleration known as the Yarkovsky effect \citep[e.g.][]{Bottke2006}, each having a 
different semimajor‑axis drift rate $da/dt$ (see Section \ref{sec3} for more details).

In total, $5\times8,668 = 43,340$ L4 clones and $5\times5,449 = 27,245$ L5 clones were numerically
integrated on separate GPUs\footnote{Two Nvidia RTX5080 in our case.} with the same setup as introduced at 
the beginning of Section~\ref{subsec21}. Every 20 days, we computed the relative distance $\Delta r$ and relative
velocity $\Delta v$ for each clone pair across all $\approx 1.9\times10^{9}$ and $\approx 7.4\times10^{8}$
configurations. We recorded all cases satisfying $\Delta r < 3\cdot10^{-3}\,\mathrm{AU}$ and 
$\Delta v < 5\,v_{\mathrm{esc}}$, where $v_{\mathrm{esc}} = \sqrt{2GM/R}$ is the escape velocity of the
corresponding hypothetical spherical parent body, providing a broader tolerance to avoid discarding 
potentially relevant candidates. The effective radius of the parent body is given by $R = (R_1^3 + R_2^3)^{1/3}$,
and its mass by $M = \tfrac{4\pi}{3} R^3 \rho$, assuming the same bulk density of 
$\rho \approx 1.5$ g cm$^{-3}$ for both components \citep[e.g.][]{Carry2012}. The individual 
diameters were derived from
\begin{equation}
D = \frac{1329}{\sqrt{p_V}}\,10^{-0.2H} [\mathrm{km}],
\label{eq:size}
\end{equation}
where $p_V$ denotes the geometric albedo and $H$ the absolute magnitude. We adopted geometric albedos from the 
\texttt{NEOWISE Diameters and Albedos V2.0}\footnote{\url{https://sbn.psi.edu/pds/resource/neowisediam.html}} 
catalogue \citep[][]{NEOWISE_DA_V2} and, in the case of missing data, we used $p_V \approx 0.075$, a value consistent 
with the observed albedo distribution \citep[e.g.][]{Grav2011a} and sufficient for our pre-selection threshold 
at this stage. The absolute magnitudes $H$ were taken from the \texttt{AstDyS} database. In total, 707 candidate 
pairs in the L4 swarm and 196 in the L5 swarm passed these criteria.

\section{Numerical simulations}\label{sec3}
We numerically integrated the orbits of all candidate pairs that passed our pre-selection procedures (see Section 
\ref{sec2}) 2.5 Myr into the past using our CUDA‑C integrator with a 1‑day time step. We generated 200 clones per
component, assigning to each clone a random Yarkovsky drift $da/dt$ from the corresponding admissible range
$[-(\mathrm{d}a/\mathrm{d}t)_{\max}, +(\mathrm{d}a/\mathrm{d}t)_{\max}]$. These clones reflect (i) 
the uncertainty in the object’s orbital solution and (ii) the uncertainty in its thermophysical parameters 
needed for modeling the perturbations caused by the Yarkovsky effect.

\subsection{Clone generation}\label{subsec13}
Here we apply the same approach as in \cite{Zizka2016}, utilizing the Cholesky decomposition. 
Let $\mathbf{e}^* = (a^*,h^*,k^*,p^*,q^*,\lambda^*)$ denote the 6D vector of nominal equinoctial 
orbital elements of the object at the given epoch (MJD 61200 in our case). This vector represents 
the best‑fit orbital solution. The initial orbital elements\footnote{Note that 
$(h,k) = e(\sin{\varpi},\cos{\varpi}), (p,q) = \tan{(i/2)}(\sin{\Omega},\cos{\Omega})$ and $\lambda = \varpi + M$.} 
of the clones sample the six-dimensional uncertainty ellipsoid around $\mathbf{e}^*$ and are given by
\begin{equation}
\boldsymbol{e} = \boldsymbol{e}^* + \boldsymbol{T}^\mathsf{T}\boldsymbol{z},
\label{elem_clones}
\end{equation}
where $\boldsymbol{z}$ is a six-dimensional vector whose components are random deviations of the standard 
normal distribution. The matrix $\boldsymbol{T}$ satisfies 
$\boldsymbol{T}^\mathsf{T}\boldsymbol{T} = \boldsymbol{\Sigma}$, where $\boldsymbol{\Sigma}$ is the 
covariance matrix of the fitted equinoctial elements provided by the \texttt{AstDyS} database, encoding both 
the formal uncertainties of the orbital elements and their mutual correlations. The matrix 
$\boldsymbol{T}$ is then obtained through the Cholesky decomposition of $\boldsymbol{\Sigma}$ 
\citep[e.g.][]{Gentle2003}.

\begin{figure}[h]
\centering
\includegraphics[width=0.9\textwidth]{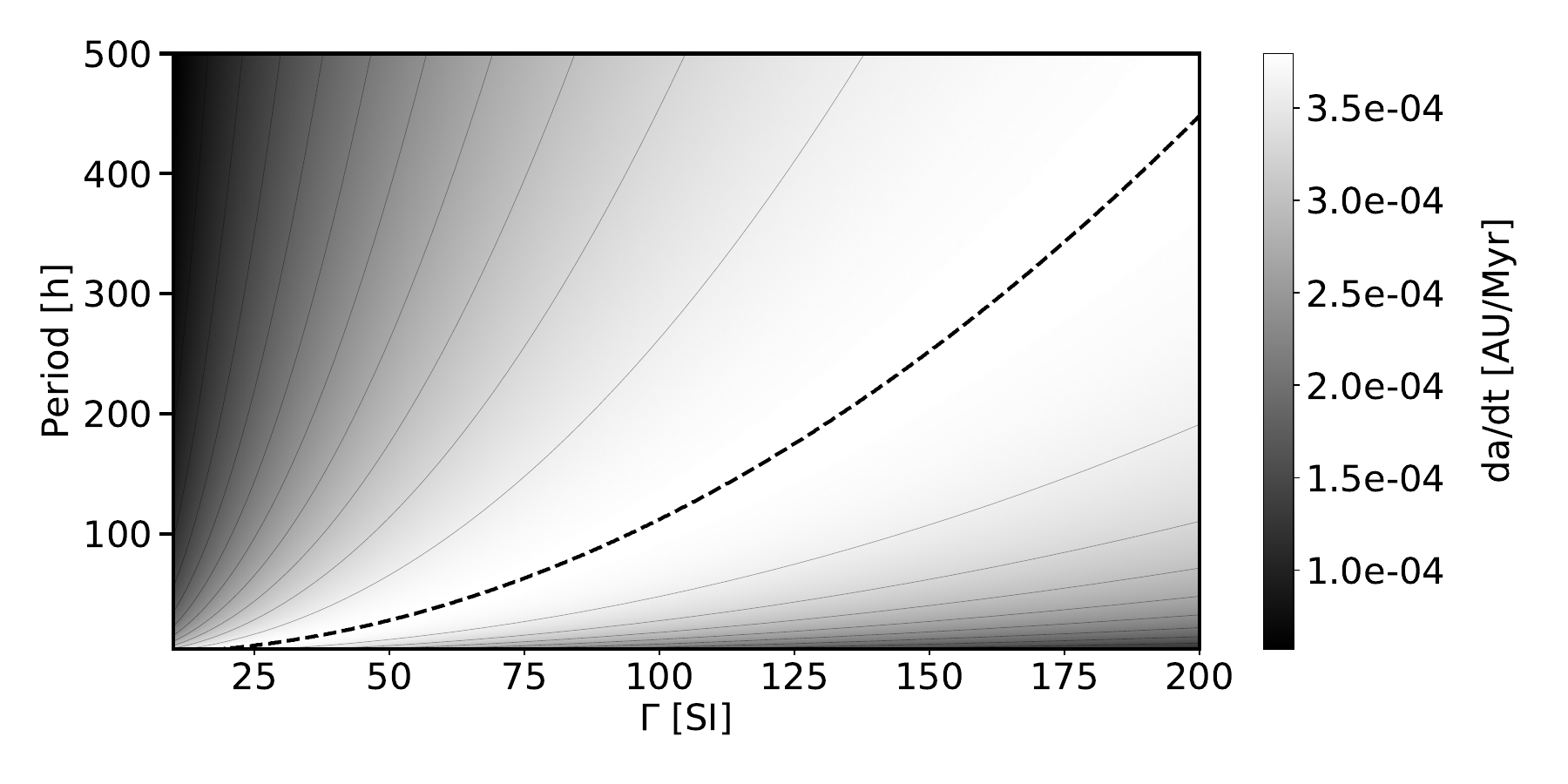}
\caption{Semimajor axis drift rates $da/dt$ for a 1 km Trojan asteroid at 5.2 AU, computed for a range of thermal
inertias and rotation periods using the linear theory of the diurnal Yarkovsky effect. The dashed line indicates
the theoretical maximum drift, $(da/dt)_{\mathrm{max}} \approx 3.79\cdot 10^{-4}~\mathrm{AU/Myr}$.}\label{fig1}
\end{figure}

To account for thermal perturbations induced by the Yarkovsky effect, we assigned each 
clone a random semimajor‑axis drift rate $da/dt$ drawn from $-(da/dt)_{\max}$ to $+(da/dt)_{\max}$, 
where $(da/dt)_{\max}$ denotes the admissible upper bound estimated for each object from the linearized 
theory of the diurnal\footnote{Because the thermal inertia $\Gamma$ of Trojans is expected below a few hundred 
SI units and their sizes are in the kilometer range, we neglect the seasonal part of the Yarkovsky effect 
in our simulation.} Yarkovsky effect presented in \cite{Vokrouhlicky1998a}. To determine $(da/dt)_{\max}$ we generated 
20,000 random realizations with thermal inertias $\Gamma \simeq 10\text{--}200~\mathrm{SI}$ and 
rotation periods $P \simeq 4\text{--}500~\mathrm{h}$ (as in \citealt{Zhuofu2025}, but slightly extended), 
while fixing the bulk density at $\rho \simeq 1.5~\mathrm{g\,cm^{-3}}$ \citep[e.g.][]{Carry2012}. 
We then computed the corresponding diurnal Yarkovsky drift $da/dt$ for a 1-km body at zero obliquity 
$\gamma$ and found $(da/dt)_{\mathrm{max}} \approx 3.79\cdot10^{-4}~\mathrm{AU/Myr}$ (see~Figure~\ref{fig1}). 
The expected maximal Yarkovsky drift, $(da/dt)_{\mathrm{max}}^{\mathrm{exp}}$, was subsequently 
determined via the simple scaling
\begin{equation}
\left(\frac{da}{dt}\right)_{\mathrm{max}}^{\mathrm{exp}} = 
\left(\frac{da}{dt}\right)_{\mathrm{max}}\cdot\frac{1}{D},
\label{scaling}
\end{equation}
where $D$ is the object's diameter, computed according to \eqref{eq:size}. At this stage, $p_V$ is taken from
\texttt{NEOWISE Diameters and Albedos V2.0} data set \citep[][]{NEOWISE_DA_V2} when available; otherwise we 
assume $p_V \approx 0.075$ \citep[c.f.][]{Grav2011a}.

Our modelling of the Yarkovsky acceleration $\boldsymbol{f}_Y$ adopts a simplified approach, retaining only the
secular semimajor‑axis drift $da/dt$ as the dominant orbital perturbation. For this purpose, we consider only the 
transverse component of the Yarkovsky acceleration to mimic the resulting $da/dt$ \citep[e.g.][]{Farnocchia2013}.

\subsection{Age estimation}\label{subsec23}
To assess whether the components of a candidate pair could have shared a common origin, a more detailed 
analysis is required. Our approach follows the strategy introduced by \cite{Vokrouhlicky2008a}, 
which also provides a method for estimating the age of the pair. For each component, we generated 200 clones, each 
assigned a plausible semimajor‑axis drift rate $da/dt$ and an initial heliocentric state vector at epoch MJD 61200. 
For every candidate pair, all clone–clone combinations were examined at five‑day intervals during the 
numerical integration, with a focus on very close encounters. We quantify the proximity between two clones 
using the Hill radius $R_{\mathrm{hll}}$, within which the components’ mutual gravity dominates over that of 
the Sun. It holds that
\begin{equation}
R_{\mathrm{hll}} = \frac{a_1 + a_2}{2}\left(\frac{M}{3M_S}\right)^{1/3},
\end{equation}
where $a_1$ and $a_2$ are the semimajor axes of the two components, $M$ is the parent‑body mass and $M_S$ 
is the solar mass.

In our CUDA implementation, each GPU processes 200 candidate pairs, with all $200 \times 200 \times 200$
clone--clone configurations evaluated at every five-day timestep. A single CUDA thread is mapped to a unique triplet 
$(i, j, k)$, where $i$ identifies the pair and $j$ and $k$ index the two clones, following the SIMT execution 
model used by NVIDIA GPUs (NVIDIA CUDA C Programming Guide). For each configuration, a fast FP32 calculation 
of the clone--clone mutual distance is compared with a predefined upper threshold that safely accounts for 
FP32 rounding. Only the accepted cases proceed to the full FP64 computations of their relative distance and velocity. 
Our setup uses eight RTX\,5080 GPUs, integrating 200 pairs on each card.

We recorded all encounter configurations at times \(t\) for which the mutual distance $\Delta r(t)$ and relative 
velocity $\Delta v(t)$ of the clones satisfied $\Delta r(t) < 5R_{\mathrm{hll}}$ and 
$\Delta v(t)~<~3v_{\mathrm{esc}}$. Candidate pairs meeting these criteria were then re-integrated with 
1500 clones per component, now requiring $\Delta r(t) < 2R_{\mathrm{hll}}$ and $\Delta v(t) < 1v_{\mathrm{esc}}$.
Those that pass even these strict conditions we regard as likely real pairs, with their convergence times~$t$ 
reflecting the distribution of the possible formation ages, as detailed in Section~\ref{sec4}. To obtain a statistically 
robust age estimate, we further require that a pair produces a sufficiently large number of such convergent 
solutions (typically $\gtrsim 1000$).

\section{Results}\label{sec4}
Applying the method described in Section~\ref{subsec23}, we require at least 1000 unique clone–clone convergent 
solutions satisfying $\Delta r < 2R_{\mathrm{hll}}$ and $\Delta v < 1v_{\mathrm{esc}}$ within the past 
2.5 Myr to obtain a reliable age estimate of the pair. We do not apply less strict conditions as is usual 
for Main Belt pairs, because the Hill spheres and escape velocities of Trojan pairs are roughly an order of 
magnitude larger, increasing the risk of interpreting a random flyby as a convergent solution of a false pair.

Under these conditions, we identified two pairs showing robust convergence even within $1R_{\mathrm{hll}}$ 
and $1v_{\mathrm{esc}}$ during the past 2.5 Myr: (264119) Georgeorton--2024~CN and (8060) Anius--(542262) 2013~BL. The orbital 
elements of their components with associated uncertainties, absolute magnitudes, and geometric albedos are listed in 
Tables~\ref{tabpairs1} and \ref{tabproperties1}.

We also found four cases involving the large bodies (1437) Diomedes and (3451) Mentor. Namely (1437) Diomedes--2015~KO417, 
(1437) Diomedes--(764465) 2013~BJ98, (3451) Mentor--2016~AO374 and (3451) Mentor--(454752) 2014~VU9. Although these cases 
show $\approx$ 100 convergent solutions, this number is far below that of our robust pairs. Moreover, the 
clone–clone encounters occur at higher relative velocities, which may point to a collisional origin or, alternatively, 
to a random flyby, warranting caution.

\begin{table}[h]
\caption{Equinoctial orbital elements of the pairs (264119) Georgeorton--2024~CN and (8060) Anius--(542262) 2013~BL, 
including their uncertainties, as of epoch MJD 61200.}\label{tabpairs1}%
\begin{tabular*}{\textwidth}{@{\extracolsep\fill}cccccccc}
\toprule
Asteroid & $a^*$ [AU] & $h^*$ & $k^*$ & $p^*$ & $q^*$ & $\lambda^*$ [$^\circ$] \\
\midrule
\textbf{264119} & 5.27845918 & 0.05959365 & 0.00678139 & 0.14753321 & 0.17864029 & 159.015119 \\
\textbf{2024~CN} & 5.13515386 & 0.06058238 & 0.00696429 & 0.16500634 & 0.16676809 & 194.581286 \\
Uncertainty & $\delta a^*$ & $\delta h^*$ & $\delta k^*$ & $\delta p^*$ & $\delta q^*$ & $\delta \lambda^*$  \\
& 1.0e-7 & 6.7e-8 & 9.4e-8 & 6.2e-8 & 8.2e-8 & 1.1e-5 \\
& 1.5e-6 & 2.9e-7 & 5.0e-7 & 1.9e-7 & 2.2e-7 & 7.1e-5 \\
& & & & & & \\
\textbf{8060} & 5.21834942 & 0.06084423 & 0.06926895 & 0.00683766 & 0.06143320 & 196.461628 \\
\textbf{542262} & 5.13462322 & -0.01141423 & 0.01793608 & 0.07186102 & 0.01753089 & 170.497521 \\
Uncertainty & $\delta a^*$ & $\delta h^*$ & $\delta k^*$ & $\delta p^*$ & $\delta q^*$ & $\delta \lambda^*$  \\
& 5.6e-8 & 3.6e-8 & 4.6e-8 & 3.2e-8 & 4.0e-8 & 6.0e-6 \\
& 1.6e-7 & 7.4e-8 & 7.4e-8 & 5.3e-8 & 6.9e-8 & 1.3e-5 \\
\botrule
\end{tabular*}
\end{table}

\begin{table}[h]
\caption{Absolute magnitudes and geometric albedos for the components of (264119) Georgeorton--2024~CN 
and (8060) Anius--(542262) 2013~BL. Absolute magnitudes are taken from the \texttt{AstDyS} database. The geometric 
albedo of (8060) Anius is listed in the \texttt{NEOWISE Diameters and Albedos~V2.0} catalogue; for all remaining 
components we adopt a default value of 0.075.}
\label{tabproperties1}%
\begin{tabular*}{\textwidth}{@{\extracolsep\fill}llcccc}
\toprule
Asteroid 1 & Asteroid 2 & $H_1$ [mag] & $H_2$ [mag] & $p_{V1}$ & $p_{V2}$ \\
\midrule
(264119) Georgeorton & 2024~CN & 13.03 & 14.82 & 0.075 & 0.075 \\
(8060) Anius & (542262) 2013~BL & 11.08 & 13.98 & 0.059 & 0.075 \\
\botrule
\end{tabular*}
\end{table}

\subsection{(264119) Georgeorton -- 2024~CN}\label{subsec14}
\begin{figure}[htp!]
\centering
\includegraphics[width=0.9\textwidth]{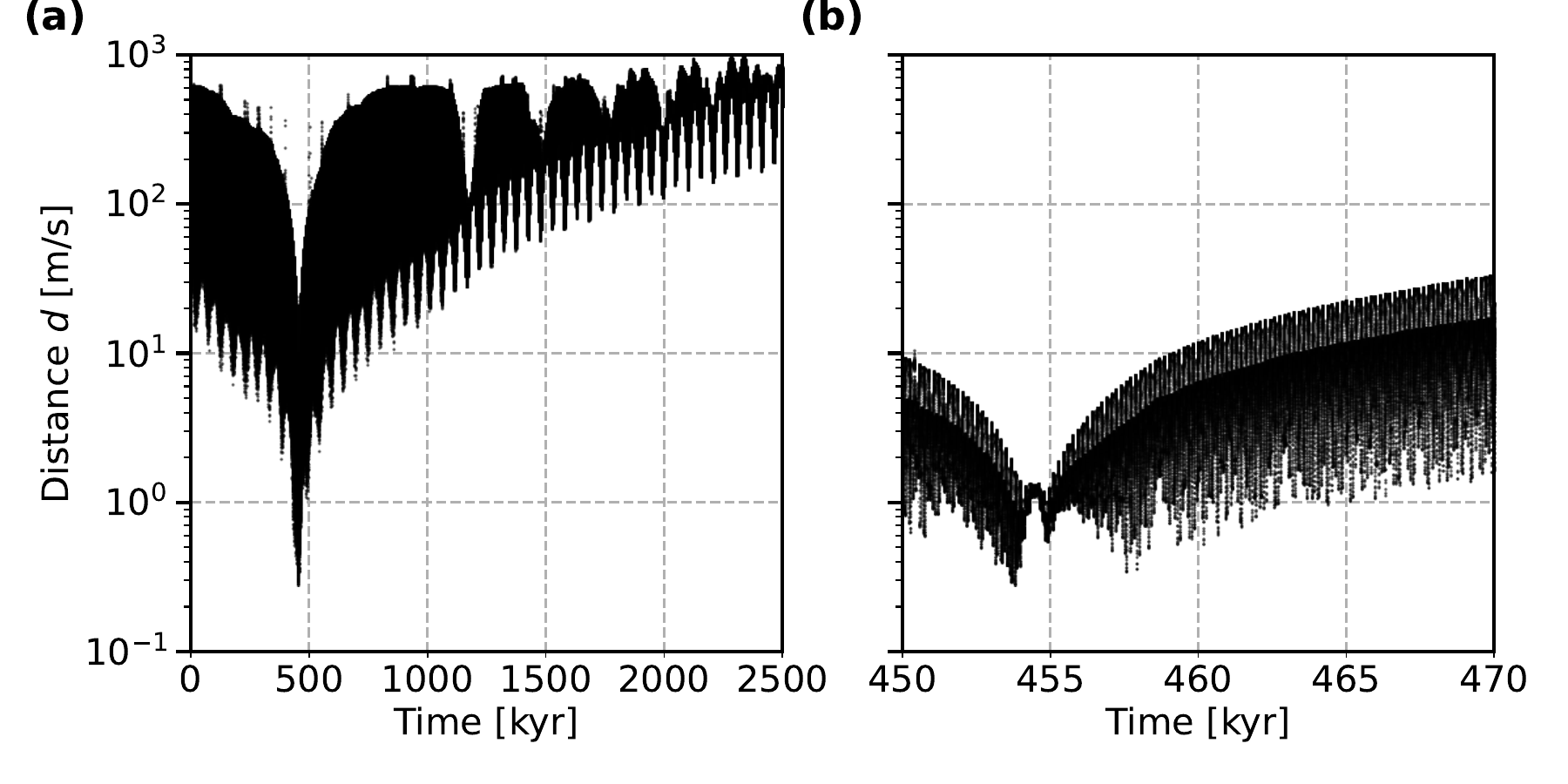}
\caption{\textit{Panel a):} The separation distance $d$ in osculating elements between the nominal orbits of the pair (264119)
Georgeorton--2024~CN (no thermal accelerations are included in this run; note the logarithmic scale of the abscissa). 
\textit{Panel b):} Close-up of the deep minimum occurring $\approx454$~kyr ago.}\label{fig2}
\end{figure}

At the reference epoch MJD 61200, the differences between the nominal osculating elements of the two components are
$\Delta a\approx1.4\times10^{-1}~\mathrm{AU}$, $\Delta e\approx1.0\times10^{-3}$, 
$\Delta i\approx3.2\times10^{-1}~\mathrm{deg}$, $\Delta \varpi\approx6.5\times10^{-2}~\mathrm{deg}$ and 
$\Delta \Omega~\approx~5.1~\mathrm{deg}$. The initial metric distance $d$ between the two nominal orbits is 
about 416 m s$^{-1}$, and it reaches an extraordinarily deep minimum of 0.28 m s$^{-1}$ approximately 454 kyr 
ago (see Figure~\ref{fig2}). We estimate the Hill radius of the parent body to be $\approx~4900~\mathrm{km}$ and 
the escape velocity from its surface $\approx 5.7$ m s$^{-1}$. 
\begin{figure}[htp!]
\centering
\includegraphics[width=0.9\textwidth]{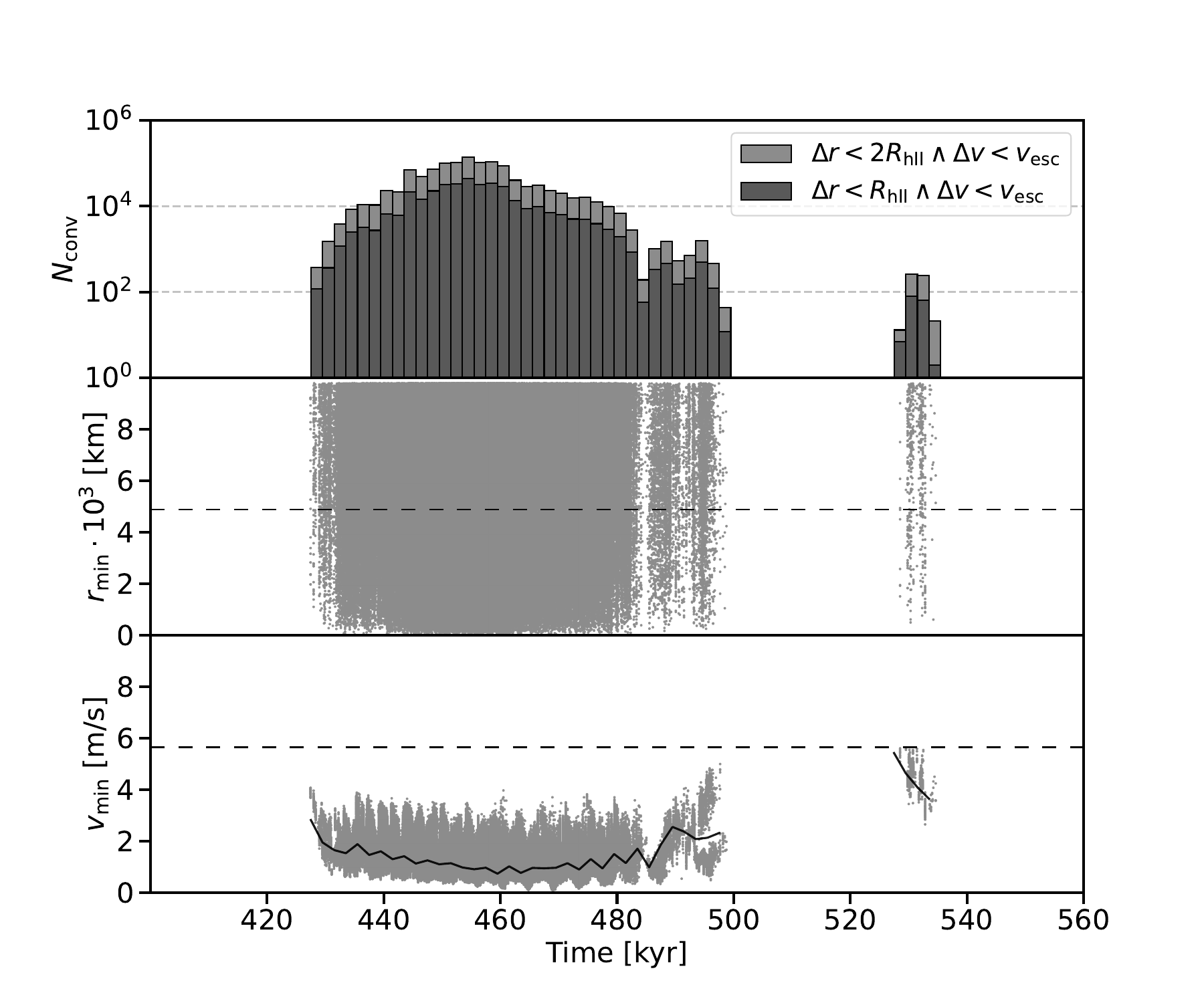}
\caption{Result of the backward integration of the pair (264119) Georgeorton--2024~CN. All quantities shown in the panels 
refer to the corresponding 2~kyr time bin. \textit{Top:} Number of unique convergent clone pairs for the two applied 
convergence criteria. \textit{Middle:} Minimum separation distance reached by each unique convergent clone pair.
\textit{Bottom:} Relative velocity of the clones at their minimum separation, together with the mean value per bin
(black line).}\label{fig3}
\end{figure}

The time distribution of convergent solutions is shown in Figure \ref{fig3}. For each solution we recorded the relative 
velocity and mutual distance of the clones. Notably, the most populated bin of the distribution roughly coincides with the 
epoch of the minimum separation distance $d$ in  osculating elements between the nominal orbits. The distribution indicates 
two possible age intervals: a dominant one at 427–500 kyr and a secondary, less populated interval at 527–536 kyr. The closest 
approaches occur at $\approx 459$ kyr ago (distance $\approx 32$ km, relative velocity $\approx 0.7$ m s$^{-1}$) in the 
first interval, and at $\approx 530$ kyr ago (distance $\approx 522$~km, relative velocity $\approx 4.6$ m s$^{-1}$) in the second. 
Across the entire simulation we found a minimum relative velocity of $\approx 0.10$~m~s$^{-1}$, reached roughly 10~kyr before 
the closest flyby, around 469~kyr ago. The very small relative velocities among converging clones suggest that the pair 
components must have separated very gently.

\subsection{(8060) Anius--(542262) 2013 BL}\label{subsec24}
\begin{figure}[htp!]
\centering
\includegraphics[width=0.9\textwidth]{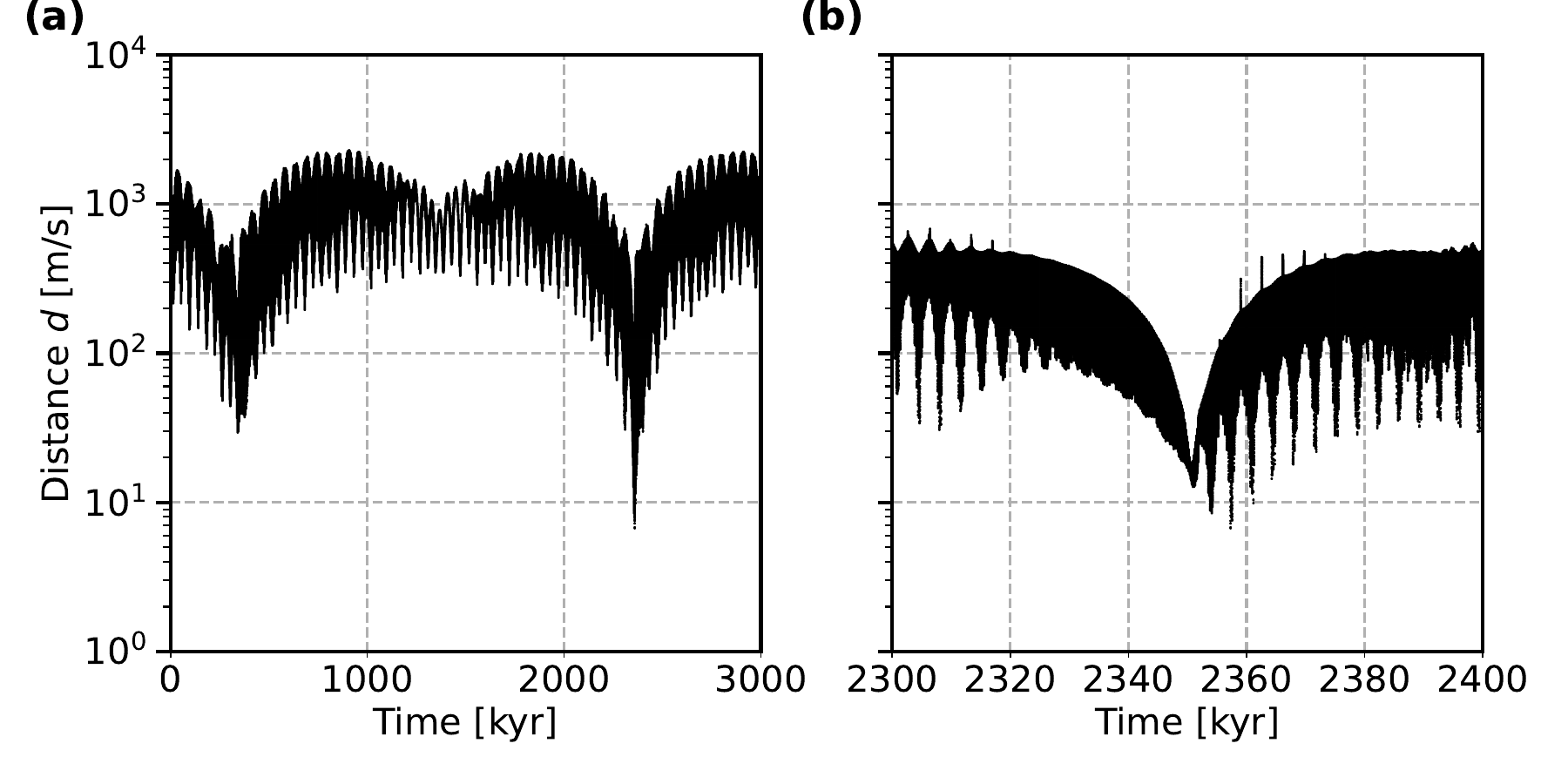}
\caption{\textit{Panel a):} The mutual distance $d$ in osculating elements between the nominal orbits of (8060)
Anius and (542262) 2013 BL (thermal accelerations neglected; note the logarithmic scale of the abscissa). 
\textit{Panel b):} Detail of the deepest minimum at $\approx 2357$~kyr ago.}\label{fig4}
\end{figure}

\begin{figure}[htp!]
\centering
\includegraphics[width=0.9\textwidth]{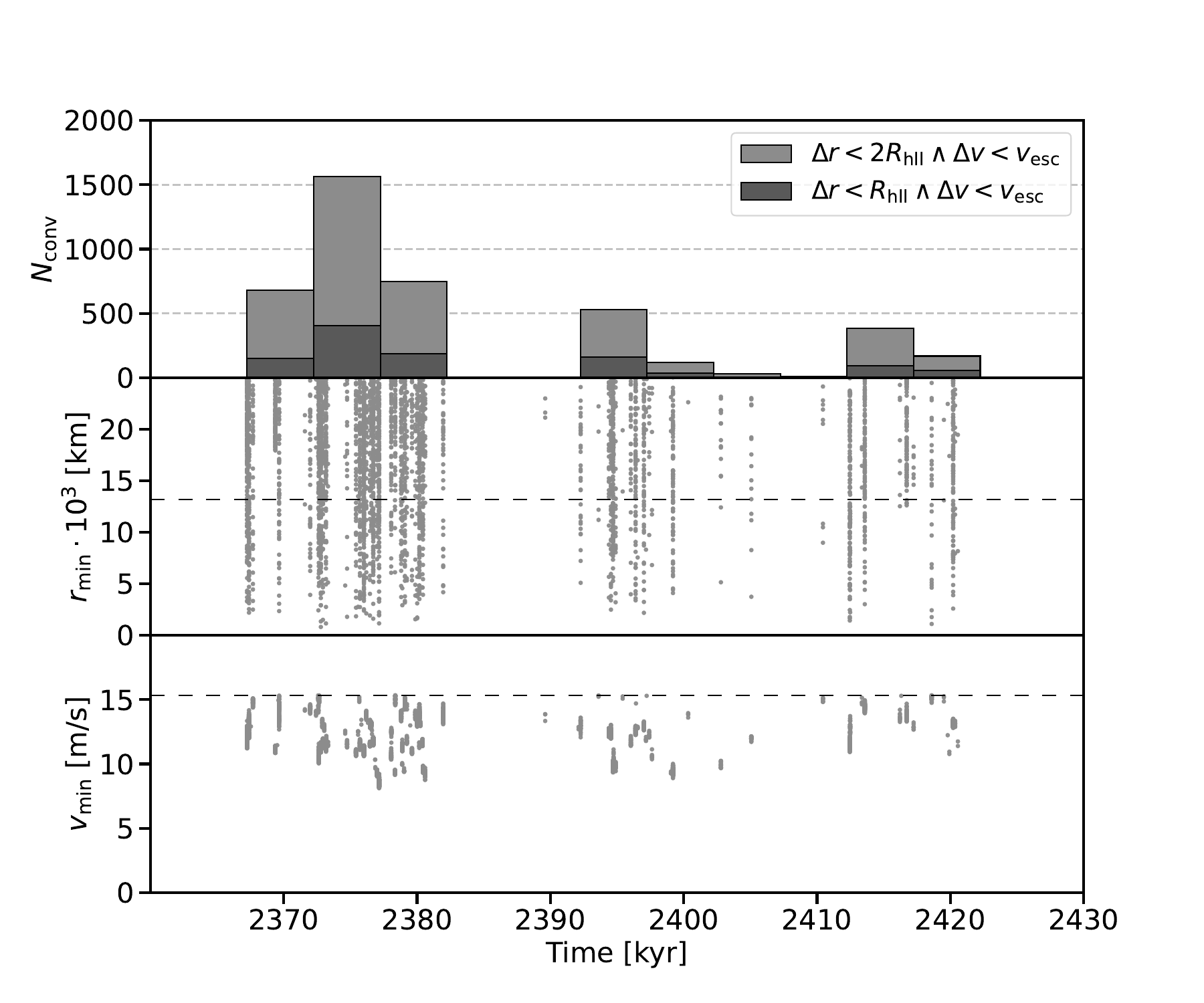}
\caption{Result of the backward integration of the pair (8060) Anius--(542262) 2013 BL. All quantities shown in the panels 
refer to the corresponding 5~kyr time bin. \textit{Top:} Number of unique convergent clone pairs for the two applied 
convergence criteria. \textit{Middle:} Minimum separation distance reached by each unique convergent clone pair.
\textit{Bottom:} Relative velocity of the clones at their minimum separation.}\label{fig5}
\end{figure}

At the reference epoch MJD 61200, the nominal osculating orbits of the two components differ notably, with 
$\Delta a \approx 8.4\times10^{-2}~\mathrm{AU}$, $\Delta e \approx 7.1\times10^{-2}$, $\Delta i \approx 1.4~\mathrm{deg}$, 
$\Delta\varpi \approx 7.4\times10^{1}~\mathrm{deg}$ and $\Delta\Omega \approx 7.0\times10^{1}~\mathrm{deg}$. The initial 
metric distance $d$ between the two nominal orbits is about 1431 m s$^{-1}$. Figure~\ref{fig4} shows that $d$ 
attains two local minima: the first one $\approx 343$ kyr ago with $d \approx 29.56$ m s$^{-1}$, and the second, deeper minimum 
$\approx 2357$ kyr ago at $d \approx 6.76$ m s$^{-1}$, which is roughly consistent with the age derived from the time distribution of 
convergent solutions (see Figure \ref{fig5}).

For the parent asteroid, the estimated Hill radius is about $13000~\mathrm{km}$, with an escape velocity of approximately
15.3 m s$^{-1}$. The closest approach among all convergent clones occurred $\approx 2373$ kyr ago at a distance 
$\approx 803$ km with a relative velocity $\approx 11.3$~m~s$^{-1}$, whereas the smallest relative velocity 
$\approx 8.1$ m s$^{-1}$ among all possible clone pairs propagated in our simulation was reached about 4 kyr earlier, 
approximately 2377~kyr~ago. The estimated age of the pair is in the range 2367–2421 kyr.

\section{Conclusion}\label{sec13}
In this work we carried out a systematic search for young asteroid pairs in the Jovian Trojans, combining large‑scale 
GPU‑accelerated backward integrations with two complementary pre‑selection strategies. We revealed two robust candidate pairs, 
(264119) Georgeorton--2024~CN and (8060) Anius--(542262) 2013 BL, showing strong and statistically significant convergence. 
Each pair was identified by both of our pre‑selection strategies. The derived formation ages of these pairs -- approximately 
427-500~kyr or 527-536~kyr for the first pair, and 2367–2421~kyr for the second -- place them among the first identified young 
asteroid pairs in the Trojan population.

The components of the pair (264119) Georgeorton--2024~CN are not linked to any known collisional family, and therefore 
we do not consider a catastrophic impact on a hypothetical parent body to be a likely formation mechanism for this system. 
The size ratio $\approx0.44$ is sufficiently below the expected ratio of $\approx 0.7$ for a vast majority of observed pairs 
in the main belt formed by YORP-induced rotational fission \citep[see][]{Pravec2019}. Combined with the very gentle separation of 
the components, this points to YORP-induced rotational fission or binary dissociation as the more probable formation mechanism.

For the second pair, (8060) Anius--(542262) 2013 BL, \cite{Marshall2022} identified its components as possible members of the 
Eurybates family. The primary, (8060) Anius, shows a somewhat less typical proper eccentricity despite having a ZTF color 
close to the family average, suggesting that (8060) Anius could also be an interloper. In our integrations, the clones of this pair 
converge within less than one Hill radius and with encounter velocities of 8-15 m s$^{-1}$ (0.5–1.0 $v_\mathrm{esc}$). 
Since the component (542262) 2013~BL is a Eurybates member, a purely interloper origin for (8060) Anius is unlikely, 
although it cannot be fully excluded. Given the large size of the primary ($\approx33$~km) and the weak solar flux 
at Jupiter Trojans, YORP‑induced rotational fission is physically improbable within the lifetime of the Solar System. 
The origin of this pair could be explained by a low‑velocity collisional breakup of a Eurybates family fragment or 
the disruption of a former binary system.
\backmatter

\bibliography{sn-bibliography}

@article{Bottke2006,
  author		= "Bottke, W. F.",
  title		= "{T}he {Y}arkovsky and {Y}{O}{R}{P} {E}ffects: {I}mplications for {A}steroid {D}ynamics",
  journal		= "Annu. Rev. Earth Planet. Sci.",
  volume		= "34",
  pages		= "157--191",
  year		= "2006",
  doi			= "10.1146/annurev.earth.34.031405.125154"
}

@article{Carry2012,
  author		= "Carry, B.",
  title		= "{D}ensity of asteroids",
  journal		= "Planet. Space Sci.",
  volume		= "73",
  pages		= "98--118",
  year		= "2012",
  doi			= "10.1016/j.pss.2012.03.009"
}

@article{Cuk2007,
  author		= "\'{C}uk, M.",
  title		= "{F}ormation and {D}estruction of {S}mall {B}inary {A}steroids",
  journal		= "Astrophys. J.",
  volume		= "659",
  pages		= "L57--L60",
  year		= "2007",
  doi			= "10.1086/516572"
}

@article{Fatka2022,
  author		= "Fatka, Petr and Moskovitz, Nicholas A and Pravec, Petr and Micheli, Marco and Devog\`ele, Maxime and Gustafsson, Annika and Kueny, Jay and Skiff, Brian and Ku\v{s}nir\'{a}k, Peter and Christensen, Eric and Ries, Judit and Brucker, Melissa and McMillan, Robert and Larsen, Jeffrey and Mastaler, Ron and Bressi, Terry",
  title		= "{R}ecent formation and likely cometary activity of near-Earth asteroid pair 2019 {P}{R}2-2019 {Q}{R}6",
  journal		= "Mon. Not. R. Astron. Soc.",
  volume		= "510",
  pages		= "6033--6049",
  year		= "2022",
  doi			= "10.1093/mnras/stab3719"
}

@article{Farnocchia2013,
  author		= "Farnocchia, D. and Chesley, S.R. and Vokrouhlický, D. and Milani, A. and Spoto, F. and Bottke, W.F.",
  title		= "{N}ear {E}arth {A}steroids with measurable {Y}arkovsky effect",
  journal		= "Icarus",
  volume		= "224",
  pages		= "1--13",
  year		= "2013",
  doi			= "10.1016/j.icarus.2013.02.004"
}

@article{Grav2011a,
  author		= "Grav, T. and Mainzer, A. K. and Bauer, J. and Masiero, J. and Spahr, T. and McMillan, R. S. and Walker, R. and Cutri, R. and Wright, E. and Eisenhardt, P. R. M. and Blauvelt, E. and DeBaun, E. and Elsbury, D. and Gautier, T. and Gomillion, S. and Hand, E. and Wilkins, A.",
  title		= "{WISE/NEOWISE} {O}bservations of the {J}ovian {T}rojans: {P}reliminary {R}esults",
  journal		= "Astrophys. J.",
  volume		= "742",
  pages		= "40",
  year		= "2011",
  doi			= "10.1088/0004-637X/742/1/40"
}

@article{Holt2020,
  author		= "Holt, Timothy R and Vokrouhlick{\'y}, David and Nesvorn{\'y}, David and Bro\v{z}, Miroslav and Horner, Jonathan",
  title		= "{A} pair of {J}ovian {T}rojans at the {L}4 {L}agrange point",
  journal		= "Mon. Not. R. Astron. Soc.",
  volume		= "499",
  pages		= "3630-3649",
  year		= "2020",
  doi			= "10.1093/mnras/staa3064"
}

@misc{Honsova1,
  author       = "{Honsov\'{a}, E.}",
  title        = "{{C}reating a catalogue of asteroid pairs and clusters}",
  howpublished = "\url{https://www.physics.muni.cz/~ehonsova/}",
  note         = "Accessed: 2026-03-15",
  year         = {2025}
}

@mastersthesis{Honsova2,
  author       = {Honsová, Eli\v{s}ka},
  title        = {{C}reating a catalogue of asteroid pairs and clusters},
  school       = {{M}asaryk {U}niversity, {F}aculty of {S}cience},
  address      = {{B}rno, {C}zech {R}epublic},
  type         = {{M}aster's thesis},
  year         = {2025},
  url          = {https://is.muni.cz/th/f5nqp/},
  note         = {Accessed: 2026-03-07}
}

@misc{NEOWISE_DA_V2,
  author       = {Mainzer, A. and Bauer, J. and Cutri, R. and Grav, T. and Kramer, E. and Masiero, J. and Sonnett, S. and Wright, E.},
  title        = {{NEOWISE} {D}iameters and {A}lbedos {V}2.0},
  year         = {2019},
  note         = {Eds.},
  publisher    = {NASA Planetary Data System},
  doi          = {10.26033/18S3-2Z54}
}

@article{Kyrylenko2021,
  author		= "Kyrylenko, I. and Krugly, Yu. N. and Golubov, O.",
  title		= "{A}steroid pairs: method validation and new candidates",
  journal		= "Astron. Astroph.",
  volume		= "655",
  pages		= "A14",
  year		= "2021",
  doi			= "10.1051/0004-6361/202140365"
}

@article{Kyrylenko2024,
  author		= "Kyrylenko, I. and Krugly, Yu. N. and Golubov, O.",
  title		= "{A}steroid pairs: {S}urvey of the inner main belt",
  journal		= "Astron. Astroph.",
  volume		= "689",
  pages		= "A291",
  year		= "2024",
  doi			= "10.1051/0004-6361/202450725"
}

@article{Levison1994,
  author		= "Harold F. Levison and Martin J. Duncan",
  title		= "{T}he {L}ong-{T}erm {D}ynamical {B}ehavior of {S}hort-{P}eriod {C}omets",
  journal		= "Icarus",
  volume		= "108",
  pages		= "18--36",
  year		= "1994",
  doi			= "doi.org/10.1006/icar.1994.1039"
}

@article{Marshall2022,
  author		= "Marschall, Raphael and Nesvorn\'{y}, David and Deienno, Rogerio and Wong, Ian and Levison, Harold F. and Bottke, William F.",
  title		= "{I}mplications for the {C}ollisional {S}trength of {J}upiter {T}rojans from the {E}urybates {F}amily",
  journal		= "Astron. J.",
  volume		= "164",
  pages		= "167",
  year		= "2022",
  doi			= "10.3847/1538-3881/ac8d6b"
}

@article{Nesvorny2026,
  author		= "David Nesvorn\'{y} and David Vokrouhlick\'{y} and Miroslav Bro\v{z} and Fernando V. Roig",
  title		= "{D}iscovery of 63 new young asteroid families",
  journal		= "Icarus",
  volume		= "443",
  pages		= "116768",
  year		= "2026",
  doi			= "10.1016/j.icarus.2025.116768"
}

@article{Pravec2010,
  author		= "Pravec, P. and Vokrouhlick{\'y}, D. and Polishook, D. and Scheeres, D. J. and Harris, A. W. and Gal{\'a}d, A. and Vaduvescu, O. and Pozo, F. and Barr, A. and Longa, P. and Vachier, F. and Colas, F. and Pray, D. P. and Pollock, J. and Reichart, D. and Ivarsen, K. and Haislip, J. and LaCluyze, A. and Ku\v{s}nir{\'a}k, P. and Henych, T. and Marchis, F. and Macomber, B. and Jacobson, S. A. and Krugly, Yu. N. and Sergeev, A. V. and Leroy, A.",
  title		= "{F}ormation of asteroid pairs by rotational fission",
  journal		= "Nature",
  volume		= "466",
  pages		= "1085--1088",
  year		= "2010",
  doi			= "10.1038/nature09315"
}

@article{Pravec2019,
  author		= "P. Pravec and P. Fatka and D. Vokrouhlick{\'y} and P. Scheirich and J. \v{D}urech and D.J. Scheeres and P. Ku\v{s}nir\'{a}k and K. Hornoch and A. Gal\'{a}d and D.P. Pray and Yu. N. Krugly and O. Burkhonov and Sh. A. Ehgamberdiev and J. Pollock and N. Moskovitz and A. Thirouin and J.L. Ortiz and N. Morales and M. Hus\'{a}rik and R. Ya. Inasaridze and J. Oey and D. Polishook and J. Hanu\v{s} and H. Ku\v{c}\'{a}kov\'{a} and J. Vra\v{s}til and J. Vil\'{a}gi and \v{S}. Gajdo\v{s} and L. Korno\v{s} and P. Vere\v{s} and N.M. Gaftonyuk and T. Hromakina and A.V. Sergeyev and I.G. Slyusarev and V.R. Ayvazian and W.R. Cooney and J. Gross and D. Terrell and F. Colas and F. Vachier and S. Slivan and B. Skiff and F. Marchis and K.E. Ergashev and D.-H. Kim and A. Aznar and M. Serra-Ricart and R. Behrend and R. Roy and F. Manzini and I.E. Molotov",
  title		= "{A}steroid pairs: {A} complex picture",
  journal		= "Icarus",
  volume		= "333",
  pages		= "429--463",
  year		= "2019",
  doi			= "10.1016/j.icarus.2019.05.014"
}

@article{Vokrouhlicky2008a,
  author		= "Vokrouhlick{\'y}, D. and Nesvorn{\'y}, D.",
  title		= "{P}airs of {A}steroids {P}robably of a {C}ommon {O}rigin",
  journal		= "Astron. J.",
  volume		= "136",
  pages		= "280--290",
  year		= "2008",
  doi			= "10.1088/0004-6256/136/1/280"
}

@article{Vokrouhlicky2017a,
  author		= "Vokrouhlick{\'y}, David and Pravec, Petr and \v{D}urech, Josef and Hornoch, Kamil and Ku\v{s}nir\'{a}k, Peter and Gal\'{a}d, Adri\'{a}n and Vra\v{s}til, Jan and Ku\v{c}\'{a}kov\'{a}, Hana and Pollock, Joseph T. and Ortiz, Jose Luis and Morales, Nicolas and Gaftonyuk, Ninel M. and Pray, Donald P. and Krugly, Yurij N. and Inasaridze, Raguli Ya. and Ayvazian, Vova R. and Molotov, Igor E. and Colazo, Carlos A.",
  title		= "{D}etailed {A}nalysis of the {A}steroid {P}air (6070) {R}heinland and (54827) 2001 {N}{Q}8",
  journal		= "Astron. J.",
  volume		= "153",
  pages		= "270",
  year		= "2017",
  doi			= "10.3847/1538-3881/aa72ea"
}

@article{Vokrouhlicky2022a,
  author		= "Vokrouhlick{\'y}, D. and Fatka, P. and Micheli, M. and Pravec, P. and Christensen, E. J.",
  title		= "{E}xtremely young asteroid pair (458271) 2010 {U}{M}26 and 2010 {R}{N}221",
  journal		= "Astron. Astroph.",
  volume		= "664",
  pages		= "L17",
  year		= "2022",
  doi			= "10.1051/0004-6361/202244589"
}

@article{Vokrouhlicky2024,
  author		= "Vokrouhlick{\'y}, David and Nesvorn{\'y}, David and Bro{\v{z}}, Miroslav and Bottke, William F. and Deienno, Rogerio and Fuls, Carson D. and Shelly, Frank C.",
  title		= "{O}rbital and {A}bsolute {M}agnitude {D}istribution of {J}upiter {T}rojans",
  journal		= "Astron. J.",
  volume		= "167",
  pages		= "138",
  year		= "2024",
  doi			= "10.3847/1538-3881/ad2200"
}

@article{Zappala1990a,
   author = {Zappal\`a, V. and Cellino, A. and Farinella, P. and Kne\v{z}evi\'c, Z.},
    title = "{{A}steroid families. {I} - {I}dentification by hierarchical clustering and reliability assessment}",
  journal = {Astron. J.},
     year = {1990},
   volume = {100},
    pages = {2030}
}

@article{Zizka2016,
  author		= "{\v{Z}}i\v{z}ka, J. and Gal\'{a}d, A. and Vokrouhlick\'{y}, D. and Pravec, P. and Ku\v{s}nir\'{a}k, P. and Hornoch, K.",
  title		= "Asteroids 87887 – 415992: the youngest known asteroid pair?",
  journal		= "Astron. Astroph.",
  volume		= "595",
  pages		= "A20",
  year		= "2016",
  doi			= "10.1051/0004-6361/201629290",
  sortkey     = {Zzizka}
}

@article{Zhuofu2025,
  author		= "Zhuofu (Chester) Li and Yasin A. Chowdhury and \v{Z}eljko Ivezi\'{c} and Ashish Mahabal and Ari Heinze and Lynne Jones and Mercedes S. Thompson and Eric Bellm and Mario Juri\'{c} and Andrew J. Connolly and Bryce Bolin and Frank J. Masci and Avery Wold and Reed L. Riddle and Richard G. Dekany",
  title		= "{E}stimates of rotation periods for {J}upiter {T}rojans with the {Z}wicky {T}ransient {F}acility photometric lightcurves",
  journal		= "Icarus",
  volume		= "438",
  pages		= "116609",
  year		= "2025",
  doi			= "10.1016/j.icarus.2025.116609"
}

@article{Vokrouhlicky1998a,
  author		= "Vokrouhlick\'{y}, D.",
  title		= "{D}iurnal {Y}arkovsky effect as a source of mobility of meter-sized asteroidal fragments. {I}. {L}inear theory",
  journal		= "Astron. Astroph.",
  volume		= "335",
  number		= "",
  pages		= "1093--1100",
  year		= "1998"
}

@book{Gentle2003,
  author		= "Gentle, J. E.",
  title		= "{R}andom {N}umber {G}eneration and {M}onte {C}arlo {M}ethods",
  address		= "New York",
  publisher	= "Springer",
  year		= "2003"
}

@inproceedings{Vinogradova2020,
  author    = {{V}inogradova, {T}amara {A}.},
  title     = {{F}amilies among the {H}ildas and {T}rojans},
  booktitle = {IAU Focus Meeting},
  volume    = {30},
  year      = {2020},
  pages     = {24--25},
  doi       = {10.1017/S1743921319003314}
}

\end{document}